\documentclass[12pt,preprint]{aastex}
\usepackage{amsmath, amsthm, amssymb}
\usepackage{natbib}
\usepackage{lscape}

\begin{document}

\title{Effects of the scatter in sunspot group
  tilt angles on the large-scale magnetic field at the solar surface}

\author{J. Jiang\altaffilmark{1}, R.H. Cameron\altaffilmark{2}, M. Sch\"{u}ssler\altaffilmark{2}}
\altaffiltext{1}{Key Laboratory of Solar Activity, National
Astronomical Observatories, Chinese Academy of Sciences, Beijing
100012, China}
\altaffiltext{2}{Max-Planck-Institut f\"ur Sonnensystemforschung,
               Justus-von-Liebig-Weg 3, 37077 G\"ottingen, Germany}

\email{jiejiang@nao.cas.cn}

\begin{abstract}
The tilt angles of sunspot groups represent the poloidal field
source in Babcock-Leighton-type models of the solar dynamo and are
crucial for the build-up and reversals of the polar fields in
Surface Flux Transport (SFT) simulations. The evolution of the polar
field is a consequence of Hale's polarity rules, together with the
tilt angle distribution which has a systematic component (Joy's law)
and a random component (tilt-angle scatter). We determine the
scatter using the observed tilt angle data and study the effects of
this scatter on the evolution of the solar surface field using SFT
simulations with flux input based upon the recorded sunspot groups.
The tilt angle scatter is described in our simulations by a random
component according to the observed distributions for different
ranges of sunspot group size (total umbral area). By performing
simulations with a number of different realizations of the scatter
we study the effect of the tilt angle scatter on the global magnetic
field, especially on the evolution of the axial dipole moment. The
average axial dipole moment at the end of cycle 17 (a
medium-amplitude cycle) from our simulations was 2.73G. The tilt
angle scatter leads to an uncertainty of 0.78 G (standard
deviation). We also considered cycle 14 (a weak cycle) and cycle 19
(a strong cycle) and show that the standard deviation of the axial
dipole moment is similar for all three cycles. The uncertainty
mainly results from the big sunspot groups which emerge near the
equator. In the framework of Babcock-Leighton dynamo models, the
tilt angle scatter therefore constitutes a significant random factor
in the cycle-to-cycle amplitude variability, which strongly limits
the predictability of solar activity.
\end{abstract}


\newpage
\section{Introduction}

\citet{Hale19} were the first to notice the systematic tilt of the
line joining the two polarities of a sunspot group with respect to
the East-West direction. The increase of the average tilt angle with
heliographic latitude later became known as `Joy's Law'.  Detailed
studies \citep[e.g.,][]{Howard91c, Sivaraman99, Sivaraman07,
Dasi-Espuig10, McClintockNorton13} used the records of sunspots
based on white-light photographs from the observatories at Mount
Wilson in the interval 1917--1985 \citep{Howard84} and at Kodaikanal
in the interval 1906--1987 \citep{Ravindra13}. Since these data do
not provide magnetic polarity information, the identification of the
leading (westward) and following (eastward) parts of sunspot groups
had to be based on visual inspection of the group morphology.
Studies based on the magnetograms \citep{Wang89b, Howard91a,
Howard91b, Tian99, Tian03, Stenflo12,Li12} more accurately define
the leading and following parts on the basis of the magnetic
polarities, but they are less complete in their coverage of sunspot
groups (or bipolar magnetic regions) and cover at most only 3
cycles.  Regardless of the type of data, all studies based on a
large sample of sunspot groups or bipolar regions confirm Joy's law,
i.e., a systematic increase of the average tilt angle away from the
equator. They also consistently find a large scatter of the
individual tilt angles about the mean.

A possible physical explanation of Joy's law and the scatter of the tilt
angles is suggested by simulations of rising magnetic flux tubes in the
rotating solar convective envelope \citep{DSilva93, Fan93, Caligari95,
Fisher95, Fan09, Weber13}. In these simulations, the Coriolis force
acting on the expanding flows along a buoyantly rising flux loop leads
to a latitudinal tilt of the loop that is consistent with Joy's
law. \citet{Weber13} show that the mean tilt angles (Joy's law) depend
on the strength of the magnetic field in the flux tubes but are not
significantly dependent on the total flux of the tube. The effects of
the turbulent convective flows on a rising flux tube were studied by
\citet{Fisher95}, \citet{LongcopeFisher96} and, more recently, by
\citet{Weber11,Weber13}. The simulations show that, for flux tubes with
field strengths above 30kG, convective flows introduce more scatter into
tubes with less flux, because such tubes are more susceptible to
deformation by convective flows. The inverse correlation between the
scatter in the tilt angle and the flux of active regions are consistent
with observations \citep[e.g. the observational results in][who suggest
a different intepretation]{Stenflo12}. Other possible physical
mechanisms for both Joy's law and the scatter in the tilt angles remain
to be explored. In this paper we restrict our attention to measuring the
scatter and determining its consequences on the evolution of the
large-scale magnetic field of the Sun.

The tilt angles of sunspot groups and, more generally, bipolar
magnetic regions have a considerable effect on the evolution of the
large-scale distribution of magnetic flux on the solar surface.  The
tilt corresponds to a latitudinal offset of the two polarities of a
bipolar region. As a result of Hale's polarity laws, this offset is
consistent across both hemispheres: during one half of a 22-year
magnetic cycle (during which, say, the North pole in the beginning
has negative polarity), the positive polarity of the emerging
bipolar regions is displaced northward from the negative polarity
(on average) in both hemispheres. The transport of the magnetic flux
by surface flows then leads to reversals of axial dipole moment and
polar fields. In the next magnetic
half-cycle, all polarities are reversed. In the framework of
Babcock-Leighton dynamos, it is through this process based upon the
tilt angle that toroidal field is converted to poloidal field and
the dynamo loop is closed \citep[see review by][]{Charbonneau:LRSP}.

Given its fundamental importance for the evolution of the
large-scale magnetic field and its role in Babcock-Leighton dynamo
models, systematic and random variations of the tilt angles are
important for a quantitative understanding of these processes
\citep{Jiang13b}. Using the Mount Wilson and Kodaikanal tilt angle
data, \citet{Dasi-Espuig10} found an anti-correlation between the
mean tilt angle (normalized by emergence latitude) of a given cycle
and the strength of that cycle \citep[see also][]{
McClintockNorton13}. \cite{Cameron10} included this observed
cycle-to-cycle variation in a surface flux transport simulation,
which reproduced the empirically derived time evolution of the solar
open magnetic flux and the reversal times of the polar fields
between 1913 and 1986.

Random and systematic variations of the tilt angles directly affect
the poloidal source term (akin to the $\alpha$-effect in mean-field
dynamo theory) of Babcock-Leighton-type dynamo models. This leads to
fluctuations in the amplitudes of the activity cycles and to
extended episodes of very low activity
\citep[e.g.,][]{Charbonneau00, Olemskoy13}. From Kitt Peak synoptic
magnetograms, \citet{Cameron13} found that occasionally a large
sunspot group with a large tilt angle emerges straddling the equator
(see their Figure 2). Such an event can strongly affect the reversal
and built-up of opposite-polarity polar field, and possibly causes
the weakness of the polar fields at the end of solar cycle 23
\citep{Cameron14}, within the context of the surface flux transport
model.

In the course of their extensive parameter study, \citet{Baumann04}
investigated the effects of the scatter in sunspot group tilt angles
on the total flux and on the polar field based on artificial solar
cycles. They found that the polar field varies by less than 30\%
when the standard deviation of the scatter was varied from 1$^\circ$
to 30$^\circ$. The objectives of this paper are to determine the
tilt angle scatter from observations and to investigate
quantitatively how strongly this observed scatter affects the
evolution of the large-scale magnetic field at the solar surface. In
particular, we consider the strength of the axial dipole moment
during activity minima as a measure of the large-scale field and
investigate which spot groups are most important in determining its
variation. We take cycle 17 as a reference cycle to do the
quantitative numerical investigation since cycle 17 is a cycle with
an average strength and not associated with a sudden increase or
decrease with respect to the adjacent cycles. In comparison, we also
consider the weak cycle 14 and the strong cycle 19. We use input
data from the Royal Greenwich Observatory (RGO) sunspot group
observations of real cycles instead of simulating artificial cycles,
in order to capture as much of the behaviour of the Sun as is
possible.

The paper is organized as follows. In Sec.~2, we consider the
dependence of the tilt angle scatter on sunspot group size using
observational data.  The surface flux transport model used to study
the evolution of the large-scale flux distribution is described in
Sec.~3. The results of the simulations with and without tilt angle
scatter are presented in Sec.~4. Our conclusions are given in
Sec.~5.

\section{Dependence of tilt angle scatter on sunspot group size and latitude}
\label{sec:measurement} We considered the tilt angles given in the
sunspot group records from Kodaikanal (30,476 sunspot groups) and
Mount Wilson (28,245 sunspot groups).  The sunspot groups were
binned according to their total umbral area, $A_U$, using bins of
equal logarithmic size (except for the first and last bins). The
tilt angle distributions were fit with Gaussians. For both data
sets, Table~\ref{tab:tilts_obs} gives the number of spot groups,
$N$, the mean tilt angle, $\left\langle \alpha \right\rangle$, the
standard deviation, $\sigma_\alpha$ and the corresponding standard
deviations of $\left\langle \alpha \right\rangle$ and
$\sigma_\alpha$ from the Gaussian fits, for each bin. Since the
emergence rate decreases towards larger sunspot groups, $N$
decreases with increasing umbral area. The mean tilt angles are in
good agreement with MWO white-light photograph analysis of
\citet{Howard96}, but are smaller than the tilt angles based on
magnetograms as reported by \citet{Howard96} and \citet{Stenflo12}.

\begin{figure}
\begin{center}
\includegraphics[scale=0.8]{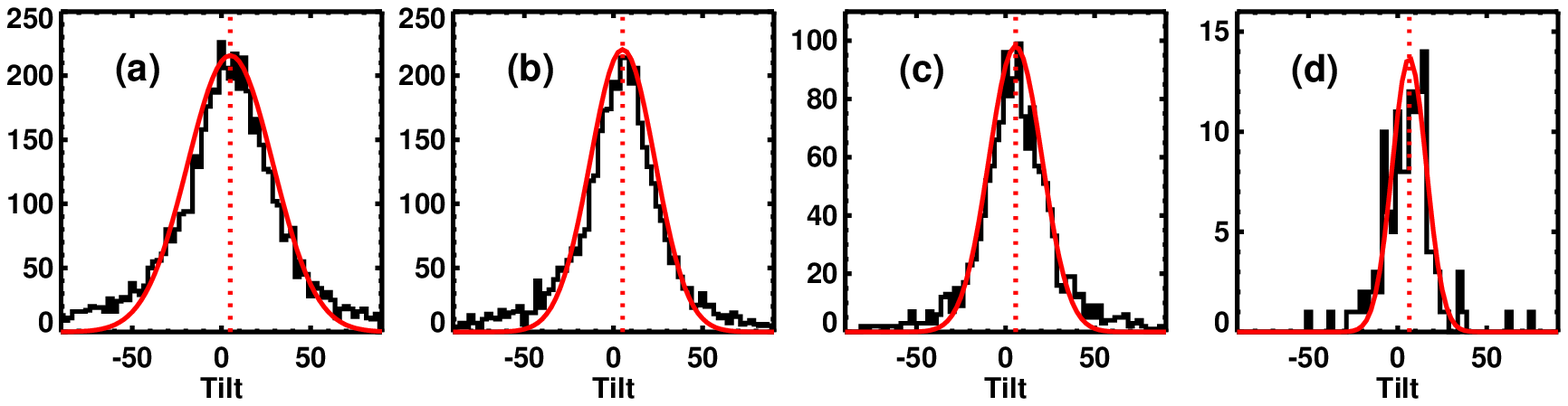}
\caption{Histograms (2.5$^\circ$ bins) of the tilt angle
distributions for different ranges of umbral area (Kodaikanal data).
Red curves show Gaussian fits. From (a) to (d), the umbra area
ranges are 10--15.8, 25.1--39.8, 63.1--100, and 158.5--251.2 in
$\mu$H, respectively.}
 \label{fig:scatter_obs}
\end{center}
\end{figure}

\begin{figure}
\begin{center}
\includegraphics[scale=0.6]{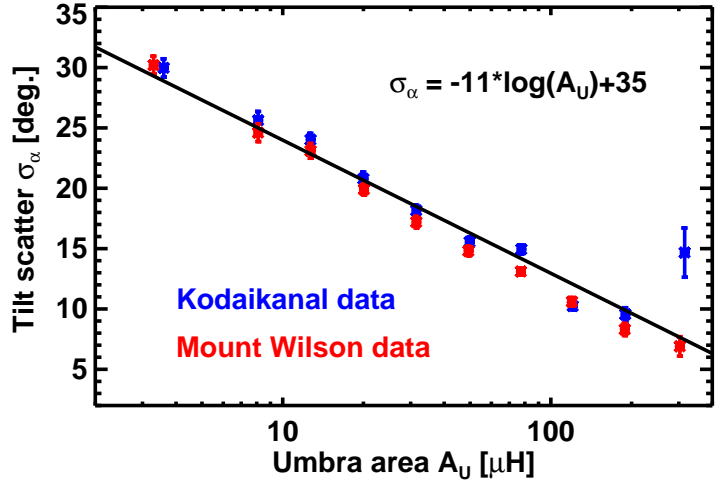}
\caption{Standard deviations of the observed tilt angle distributions
binned in logarithmic umbral area. Kodaikanal data are indicated in
blue, Mount Wilson in red. The solid curve represents a fit with the
function given by Eq.~(\ref{eq:tilt_scatter}).}
\label{fig:fit_tilts}
\end{center}
\end{figure}

Consistent with \citet{Stenflo12}, we find that the tilt angle
scatter strongly decreases with increasing sunspot group area (see
Table~\ref{tab:tilts_obs}). In contrast to \citet{Stenflo12}, we
also find that the mean tilt angle increases somewhat with sunspot
group area, but only for the Kodaikanal data, so that the relevance
of this trend is unclear. Hereafter we therefore do not consider a
size dependence of the mean tilt angle, but only its the latitudinal
dependence (Joy's Law).  Figure \ref{fig:scatter_obs} shows four of
the tilt angle distributions in $2.5^\circ$ bins from Kodaikanal
record together with the Gaussian fits (red curves) from which the
values given in Table~\ref{tab:tilts_obs} were derived. Figure
\ref{fig:fit_tilts} shows the tilt angle scatter, $\sigma_\alpha$,
as a function of umbral area, $A_U$. The error bars indicate
one-sigma error estimates. The data shown in
Figure~\ref{fig:fit_tilts} can be represented by the expression
\begin{equation}
\sigma_\alpha=-11*\log(A_U)+35. \label{eq:tilt_scatter}
\end{equation}

We also studied the latitude dependence of the scatter.
Table~\ref{tab:lat} shows that there is no significant latitudinal
dependence of the scatter on the tilt angle. These results are
consistent with the visual impression given by Figure 2 of
\citet{Fisher95} where a strong dependence of the HWHM of the
distributions with respect to area can be seen, with no obvious
dependence on latitude.

\section{Model description}
\subsection{Surface flux transport model}
The surface flux transport (SFT) model \citep[see the review
by][]{Mackay12} describes the evolution of the large-scale magnetic flux
distribution at the solar surface as a combined result of the emergence
of bipolar magnetic regions, a random walk due to supergranular flows,
and the transport by large-scale surface flows
\citep[e.g.,][]{Wang89,Ballegooijen98,Mackay02,Schrijver02,Baumann04}.
The relevant equation is
\begin{eqnarray}
\frac{\partial B}{\partial t}=&-&\omega(\theta)\frac{\partial
B}{\partial \phi}-\frac{1}{R_\odot\sin\theta}
\frac{\partial}{\partial\theta}\left[\upsilon(\theta)B\sin\theta\right]\nonumber \\
&+&\frac{\eta}{R_\odot^{2}}\left[\frac{1}{\sin\theta}\frac{\partial}{\partial\theta}
\left(\sin\theta\frac{\partial B}{\partial\theta}\right)+
\frac{1}{\sin^{2}\theta}\frac{\partial^{2}B}{\partial\phi^2}\right]\nonumber\\
&+&S(\theta,\phi,t),
\end{eqnarray}
where $B$ is the radial component of the magnetic field, $\theta$ is
the heliographic colatitude ($\lambda=\pi/2-\theta$ is the
latitude), and $\phi$ is the heliographic longitude. For the surface
differential rotation, $\omega(\theta)$, we use the profile given by
Snodgrass (1983)
$\omega(\theta)=13.38-2.30\cos^2\theta-1.62\cos^4\theta$-13.2 (in
deg day $^{-1}$). The surface meridional flow, $\upsilon(\theta)$,
is described by the profile suggested by van Ballegooijen et al.
(1998), i.e.,
\begin{equation}
\upsilon(\lambda)=\left\{
  \begin{array}{l l}
     \upsilon_0\sin(180^\circ\lambda/\lambda_0) & |\lambda| \leq \lambda_0 \\
     0 & \textrm{otherwise},
  \end{array}
  \right.
\end{equation}
with $\upsilon_0=11$~ms$^{-1}$ and $\lambda_0 = 75^\circ$.  The
turbulent diffusivity that models the random walk of magnetic
features associated with supergranulation is taken as
$\eta=250\,$km$^2$s$^{-1}$. This value is in the middle of the range
given by \cite{Schrijver00} \citep[see also][]{Jiang13b}.  The
source of the magnetic flux, $S(\theta,\phi,t)$, describing the
emergence of the sunspot groups is discussed in the subsequent
section.

For the numerical simulations we used the code originally developed
by  \citet{Baumann04}. The magnetic field is expressed in terms of
spherical harmonics up to  $l$ = 63. A fourth-order Runge-Kutta
method is used for time stepping.

\subsection{Sources of magnetic flux}
\label{sec:source} We follow the method described by
\citet{Cameron10} and \citet{Jiang11b} to generate the source term,
$S(\theta,\phi,t)$, from the RGO sunspot record. In brief, each
observed sunspot group is regarded as a bipolar magnetic region
(BMR) with modified Gaussian distributions for the preceding and
following parts \citep{Baumann04}. Polarities are chosen according
to Hale's laws, taken into account the cycle overlap around activity
minima. Each BMR is introduced into the SFT simulation at the time
of maximum sunspot area of the corresponding sunspot group.

We use the RGO record since it is by far the longest and most
complete in terms of coverage of sunspot groups. The disadvantage of
RGO data, however, is the absence of information concerning the tilt
angles. We therefore assign the tilt angle for each BMR according to
the relation determined by \citet{Jiang11a},
\begin{equation}
\alpha=0.7\, T_n\sqrt{|\lambda|}+\epsilon,
\label{eq:tilt}
\end{equation}
where $T_n$ ($n$ is the cycle number) represents the systematic
variation of the mean tilt angle between solar cycles, as determined
by \citet{Jiang11a} (see their Figure~11 and Equation~15). In the
cycles studied here, we have $T_{14}=1.5$, $T_{17}=1.3$ and
$T_{19}=1.0$. Combining the RGO sunspot area records with $T_n$ for
different cycles, the study by \citet{Cameron10} can be extended
back to 1874. The factor 0.7 was introduced and calibrated by
\citet{Cameron10} to reproduce the observed ratio between the maxima
and minima of the open heliospheric flux. It possibly results from
the reduction of the tilt angles by near-surface inflows towards
sunspot groups \citep{Gizon04,Jiang10}.

The scatter of the tilt angles is modeled by $\epsilon$ in
Equation~(4), which is drawn from random distributions consistent
with the observationally inferred standard deviations in the various
$A_U$ bins (cf. Table~\ref{tab:tilts_obs}).  We assume $\epsilon$
for each BMR to be an independent realization of a random process
with a Gaussian distribution with zero mean and a half width
according to the relationship between scatter and umbral area given
by Equation~(1).

We first study cycle 17 (1933.8-1944.2), which has a medium cycle
amplitude during the period of RGO data, with a maximum $R_g=123$ of
the 12-month running mean of the group sunspot number. We discuss
the results for cycle 19 (the strongest cycle during the period
covered by the RGO data) and cycle 14 (the weakest cycle) in Section
\ref{sec:dependence_cycles}. As the initial condition for the SFT
simulations, we take the field distribution at the time 1933.8 from
the extended simulation of \citet{Cameron10}, where we use the
values of $T_n$ according to its relationship to cycle strength
\citep[in the form given by][] {Jiang11a}. Since it takes roughly 2
years for low-latitude magnetic flux to be transported to the poles,
we run the SFT simulations until 2 years after the minimum between
cycles 17 and 18, i.e., until 1946.2. Similar procedures are
followed for cycles 14 and 19.

\subsection{Averaged quantities}
\label{sec:quantities} The output of the SFT simulations is the
radial component of the magnetic field at the solar surface as a
function of colatitude, longitude and time, $B(\theta,\phi,t)$. A
more compact representation of the results is obtained by
considering the longitudinally averaged field as a function of
colatitude and time,
\begin{equation}
\left\langle B
\right\rangle(\theta,t)=\frac{1}{2\pi}\int_0^{360}B(\theta,\phi,t)d\phi,
\label{eq:bavg}
\end{equation}
which yields a ``magnetic butterfly diagram''.  Several one-dimensional
time series can be constructed from $\left\langle B
\right\rangle(\theta,t)$ by performing weighted integrals over various
(co)latitude ranges. Among these are the polar fields of each
hemisphere, which we here define as the average (signed) field poleward
of $\pm75^\circ$ latitude in each hemisphere. For the north polar field
we thus write
\begin{equation}
  B_{\rm NP} = \int_0^{15} \left\langle B\right\rangle(\theta,t)
              \sin\theta d\theta\left/\int_{0}^{15}\sin\theta d\theta\right.,
\label{eq:polar}
\end{equation}
and analogous for the south polar field, $B_{\rm SP}$.
Other relevant quantities are the  axial dipole moment,
\begin{equation}
D(t)=\frac{3}{2} \int_0^{180} \left\langle B\right\rangle(\theta,t)
              \cos\theta\sin\theta d\theta,
\end{equation}
\label{eq:dipole}
the low-latitude contribution to the dipole moment,
\begin{equation}
D_{<55}(t)=\frac{3}{2} \int_{35}^{145}  \left\langle B\right\rangle(\theta,t)
              \cos\theta\sin\theta  d\theta,
\end{equation}
\label{eq:dipole55}
and the quantity
\begin{equation}
S_{<55}(t)=\int_{35}^{145} |\left\langle B\right\rangle(\theta,t)|
              \sin\theta d\theta\left/\int_{35}^{145}\sin\theta d\theta\right.,
\label{eq:bavg55}
\end{equation}
which we introduce as a measure of the structure in the mid- and
low-latitude part of the magnetic butterfly diagram.

\section{Results}

We study the effect of the scatter in the tilt angles on the
evolution of the Sun's large-scale magnetic field by comparing
simulations with and without the scatter.

\subsection{Evolution without tilt angle scatter}
\label{sec:noscatter}

Figure \ref{fig:noscatter} shows results of the simulation for cycle
17 with no scatter in the tilt angles, i.e., $\epsilon=0$ in
Equation~(\ref{eq:tilt}).  Panel (a) shows the magnetic butterfly
diagram, i.e., the time-latitude plot of $\langle B \rangle$.
Poleward surges of magnetic flux in both hemispheres illustrate the
transport magnetic flux with following polarity (opposite to the
polarity of the polar field during rise phase of the cycle) from the
activity belt to the poles. They reverse the old polar field of
cycle 16 and build up the polar field of cycle 17. The time
evolution of $S_{<55}(t)$, which represents the amount of structure
in the magnetic butterfly diagram at mid and low latitudes, is given
in panel (b) along with the observed sunspot number (in red). The
value of $S_{<55}(t)$ at a given time is mainly determined by the
product of area and modulus of the tilt of the BMRs present. Its
evolution roughly follows the solar cycle and peaks around the cycle
maximum in 1937.9.  Panel (c) shows the evolution of the total axial
dipole moment, $D$ (black curve), and the contribution of the
low-latitude flux, $D_{<55}$ (red curve). During the initial phase
of the simulation, the positive north polar field and negative south
polar field correspond to a positive dipole moment. BMRs emerging in
the course of cycle 17 then contribute negative dipole moments at
low latitude as seen in $D_{<55}$ (the positive values during the
beginning of the cycle are due to cycle overlap). With the
subsequent transport towards the poles, the global dipole moment
decreases and reverses around 1938.5. It peaks just after the cycle
minimum (1944.6). Panel (d) shows the corresponding evolution of the
polar fields, which reach their (unsigned) maxima about 2 years
after solar minimum. Note that the timing of the polar field
reversals depends on the definition of the `polar cap' and on
whether the radial component or the line-of-sight component with
respect to the ecliptic of the magnetic field is considered. The
difference can amount up to two years \citep{Jiang13a}.

The second column in Table~\ref{tab:comparison} gives numerical values
for the maxima of the quantities discussed above. The asymmetry of the
polar fields is due to the hemispherically asymmetric sunspot emergence.

\subsection{Evolution with tilt angle scatter}

We evaluated the effects of the tilt angle scatter by performing SFT
simulations of cycle 17 for which each emerging BMR was associated
with a random perturbation of the tilt angle, $\epsilon$, according
to Equation~(\ref{eq:tilt}). The values for $\epsilon$ were taken
from a Gaussian distribution with zero mean and a standard deviation
$\sigma_\alpha$ based on Equation~(\ref{eq:tilt_scatter}). We
carried out 50 simulation runs with different realizations of
$\epsilon$. The standard deviations, associated with 50 random
realizations, of the quantities defined in
Section~\ref{sec:quantities} reflect the effects of the scatter in
sunspot group tilt angles on the large-scale field.

\begin{figure}
\begin{center}
\includegraphics[scale=1.0]{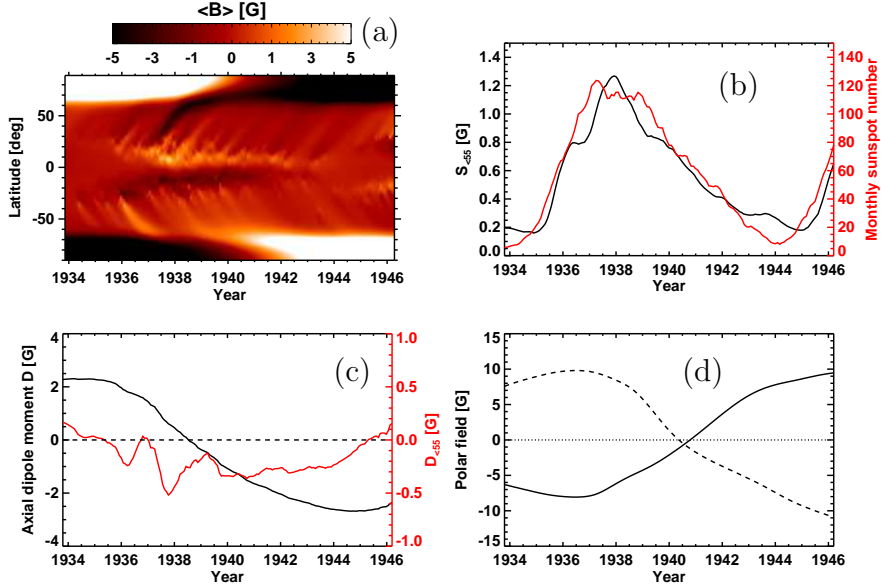}
\caption{Evolution of the simulated large-scale magnetic field during
cycle 17 for the case without tilt angle scatter. (a): time-latitude
diagram of $\langle B \rangle$; (b): low-latitude structure, $S_{<55}$
(black), and observed sunspot number (red); (c): total axial dipole
moment, $D$ (black), and contribution to $D$ of the latitudes below
$\pm55^\circ$ (red); (d): North (dashed) and South (solid) polar
field. For definitions of these quantities, see
Eqs.~(\ref{eq:bavg}) to (\ref{eq:bavg55}).}
\label{fig:noscatter}
\end{center}
\end{figure}

\begin{figure}
\begin{center}
\includegraphics[scale=1.0]{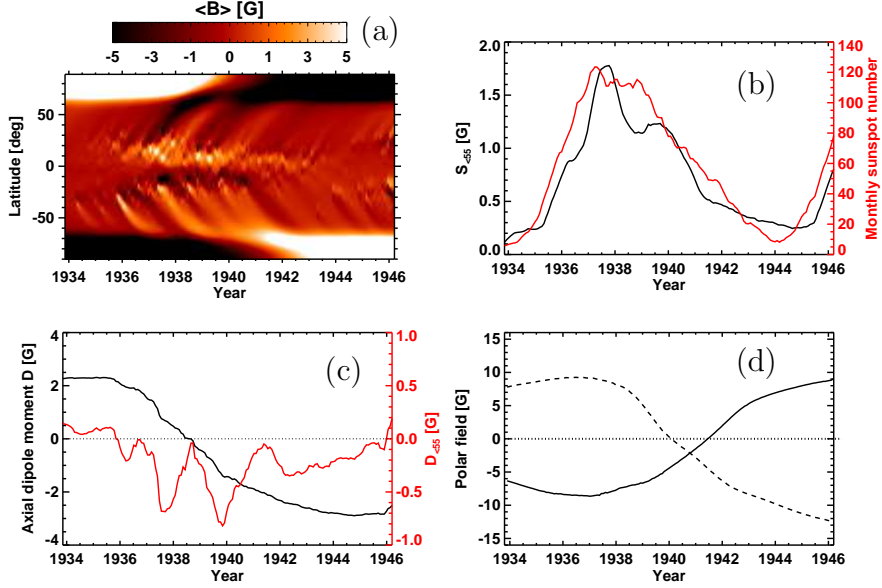}
\caption{Similar to Figure~\ref{fig:noscatter}, but for one simulation run
 (out of 50) with random scatter of the tilt angles according to the
 observed distributions. This example develops a similar dipole field as
 the  case without the tilt angle scatter.}
\label{fig:with_tilt2}
\end{center}
\end{figure}

Figure \ref{fig:with_tilt2} illustrates the result for one of these
runs. The simulated magnetic butterfly diagram, shown in panel (a),
appears more `grainy' than the corresponding plot in
Figure~\ref{fig:noscatter}, which is the result of some BMRs
emerging with randomly occurring large tilt angles. This graininess
is represented by an increase of the quantity $S_{<55}$ by about
40\% compared to the case without tilt angle scatter (cf. panel (b)
in Figure \ref{fig:noscatter}). In the case of SFT simulations
without tilt angle scatter, the ratio of the net magnetic fluxes in
the activity belts and in the polar regions is usually lower than in
the observations.  Examples are Figure 6 of \citet{Schuessler06} and
Figure 3 of \citet{Yeates14}. The tilt angle scatter increases the
averaged net flux at the low latitudes without increasing the net
flux at high latitudes. There are also more poleward surges of
opposite-polarity flux. The magnetic butterfly diagrams for
simulations with tilt angle scatter are therefore more similar to
their observed counterpart for the last 3 cycles \citep[see,
e.g.,][]{Hathaway10}.

\begin{figure}
\begin{center}
\includegraphics[scale=1.0]{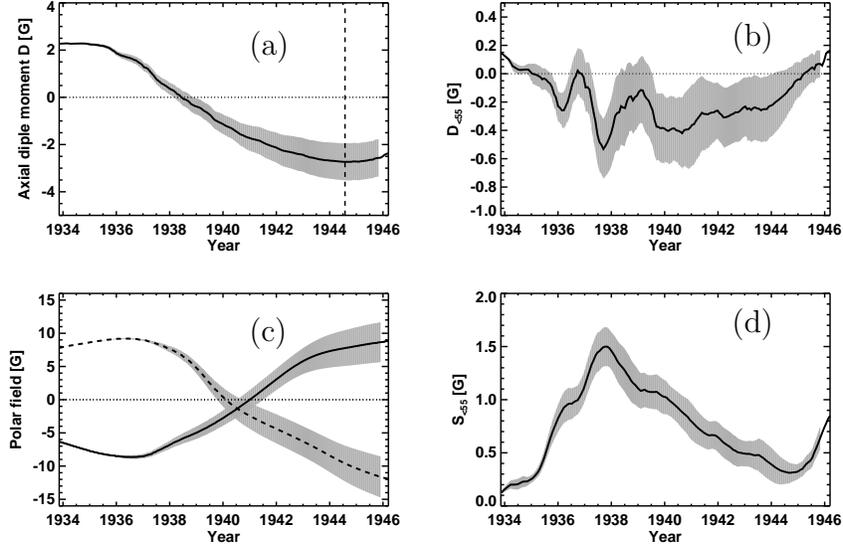}
\caption{Evolution of various quantities for simulations including
  random scatter of the tilt angles. Solid curves show averages of 50
  simulations with different sets of random numbers while gray shades
  indicate the corresponding standard deviations.  (a): total axial
  dipole moment, $D$; (b): contribution to $D$ of the latitudes below
  $\pm55^\circ$; (c): North (dashed) and South (solid) polar field; (d):
  mean net flux density within $\pm55^\circ$ latitude. For
  definitions of these quantities, see Eqs.~(\ref{eq:polar}) to
  (\ref{eq:bavg55}).}
  \label{fig:with_tilt1}
\end{center}
\end{figure}

For the case shown in Figure~\ref{fig:with_tilt2}, the dipole field
in panel (c) and the polar field in panel (d) exhibit a similar time
evolution as in the case without the tilt scatter.  However, this is
not a general feature as can be seen in Figure~\ref{fig:with_tilt1},
which shows the averages (solid and dashed curves) and the standard
deviations (grey shades) of these quantities for the 50 SFT
simulations with tilt angle scatter. Panel (a) gives the time
evolution of the axial dipole moment, $D$. The average curve is
close to the case without the tilt angle scatter shown in
Figure~\ref{fig:noscatter}. Since the dipole moment is built up from
the accumulated contributions of emerging BMRs, the standard
deviation increases with time. It starts at zero since the initial
condition was the same for all runs.  At the time of maximum dipole
moment (indicated by the dashed vertical line), the standard
deviation due to the random scatter of the tilt angles amounts to
0.78G, which is about 30\% of the average value at the end of cycle
17. The relative variation (30\%) depends on the mean dipole moment
at the end of the cycle \citep[which is correlated with the strength
of the following cycle,][]{Jiang07}.

The contribution of the lower latitudes to the axial dipole moment,
$D_{<55}$, shown in panel (b) of Figure~\ref{fig:with_tilt1} is
strongly affected by the tilt angle scatter. Finally, panel (d)
shows the quantity $S_{<55}(t)$, which represents the amount of
structure in the magnetic butterfly diagram. The maximum of its
average time profile is about 20\% higher than for the case without
the tilt scatter.  This results from strongly tilted BMRs, which
occasionally appear in the runs with scatter and contribute
significantly to $S_{<55}(t)$. A comparison between the values for
the various quantities discussed above for the cases with and
without scatter is given in Table~\ref{tab:comparison}.

\subsection{Dependence on sunspot group size}

In order to study the dependence of the effect of tilt angle scatter on
group size, we divided the sunspot groups of cycle 17 into 5 samples of
approximately equal total umbral area.  We then carried out $5\times50$
SFT simulations for which the random component of the tilt angle,
$\epsilon$, was only introduced in one of these samples while the others
had $\epsilon=0$.

Table \ref{tab:diff_size} summarizes the results of these
simulations, where $N$ denotes the number of the sunspot groups in
each bin. We see that the contribution of each bin to the standard
deviation of the axial dipole moment scales roughly as $1/\sqrt{N}$,
so that the effect of the tilt angle scatter decreases
systematically for the (more numerous) small groups. We therefore
expect that the effect of the scatter in the tilt angles of the
abundant ephemeral regions is negligible.

\subsection{Dependence on sunspot group latitude}
In order to study how the effect of the tilt angle scatter depends
on the latitude of emerging BMRs, we divided the sunspot groups of
cycle 17 in 5 latitude bins. We then carried out $5\times 50$ SFT
simulations for which the tilt angle scatter was only applied to the
groups belonging to one of these bins.  The bins and the results of
these simulations are shown in Table \ref{tab:diff_lati}.

Similar to the size dependence discussed in the preceding
subsection, we find that the averages of the maximum values of the
dipole moment and of the polar fields are almost unaffected
(cf.~Table~\ref{tab:comparison}). However, the latitude dependence
of the standard deviations of the axial dipole and the polar fields,
i.e., their uncertainty due to the random component, is quite
strong: sunspot groups emerging below 15$^\circ$ contribute much
more strongly to the uncertainties than those appearing in higher
latitudes. As explained in the above subsection, the uncertainties
result from a combination of two factors: the number of sunspot
groups $N$ in a given latitude bin and their individual contribution
to the dipole and polar fields. Although there are 3 times less
sunspot groups in the 0$^\circ$-5$^\circ$ bin than that in the
10$^\circ$-15$^\circ$ bin, both latitude ranges contribute similarly
to the standard deviation of the dipole moment. This reflects the
fact that individual sunspot groups at lower latitudes affect the
evolution of the dipole field more strongly.

To illustrate the strong dependence of the uncertainty on emergence
latitude, we performed SFT simulations with single BMRs initially
placed at different latitudes. We chose a sunspot group with
$A_U=1000$~$\mu$H, corresponding to a total magnetic flux of
6$\times10^{21}$Mx, and assumed a large tilt angle of 80$^\circ$. We
considered cases where the BMR was inserted at latitudes between
0$^\circ$ and 40$^\circ$, in steps of 10$^\circ$.

\begin{figure}
\begin{center}
\includegraphics[scale=0.42]{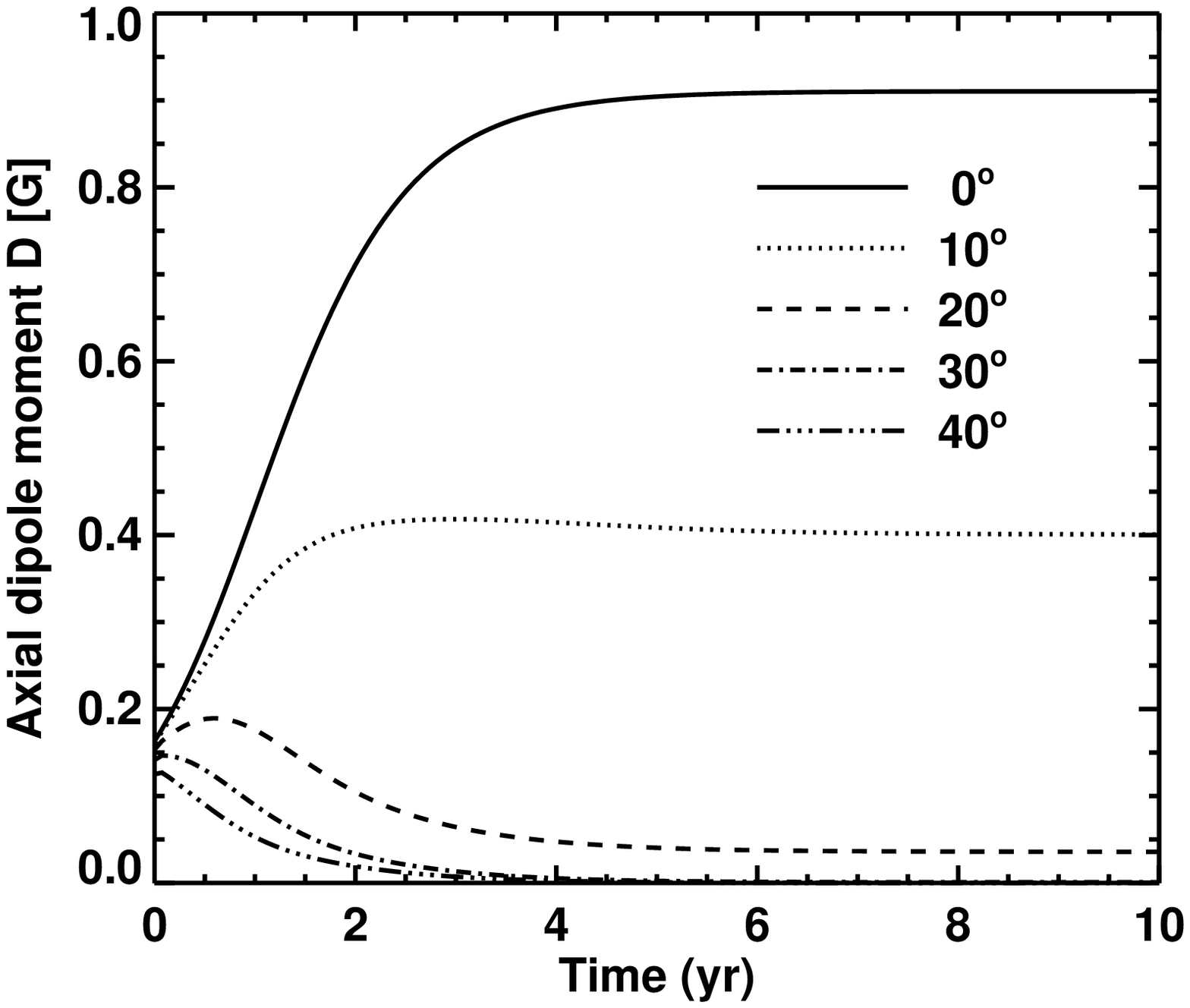}
\includegraphics[scale=0.42]{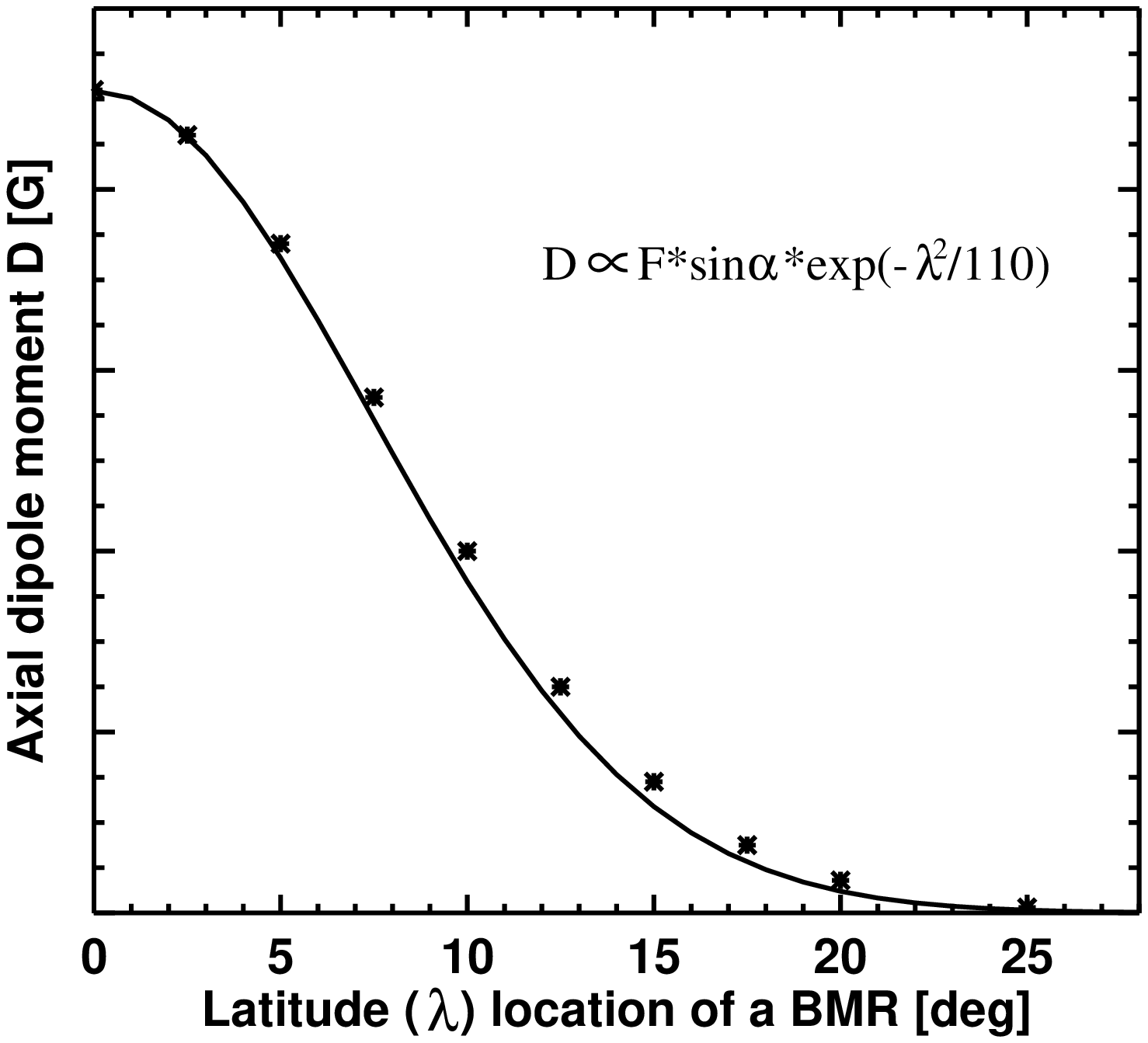}
 \caption{Left panel: time evolution of the axial dipole moment
resulting from a single BMR with a total flux of 6$\times10^{21}$ Mx
and tilt angle 80$^\circ$, for various emergence latitudes; Right
panel: relation between the eventual equilibrium axial dipole moment
and the latitudinal location of a BMR with given flux, F, and tilt
angle, $\alpha$. The points are the results from SFT simulations for
different BMRs emerging at different latitudes. The solid curve is
the Gaussian fit of the form $\exp(-\lambda^2/110)$.}
\label{fig:single_BMR_dip}
\end{center}
\end{figure}

The left panel of Figure \ref{fig:single_BMR_dip} shows the
evolution of the axial dipole moment for the various emergence
latitudes.  For cross-equator emergence (0$^\circ$), the centroids
of the two polarities are initially located at about $\pm4.3^\circ$
latitude. Advection by the meridional flow in each hemisphere
separates the polarities and increases the dipole moment. At the
same time, about half of the magnetic flux diffuses across the
equator, where it cancels with opposite-polarity flux.  The
remaining flux is eventually transported to the poles and the dipole
moment reaches a plateau of 0.9~G, corresponding to polar fields of
$\pm3.7$~G. These values represent about 20\% of the simulated
dipole and polar fields generated by all recorded sunspot groups of
cycle 17. At the other extreme, the axial dipole moments due to BMRs
emerging at 30$^\circ$ or 40$^\circ$ steadily decay as both
polarities are swept together towards the pole and cancel there,
while only a negligible amount of magnetic flux is transported
across the equator. The intermediate cases of emergence at
10$^\circ$ and 20$^\circ$, show a mixture of the two types of
behavior. The right panel of Figure \ref{fig:single_BMR_dip} shows
the relation between the final axial dipole moment and the
latitudinal location of the BMR with a given magnetic flux and tilt
angle. The behaviour corresponds to a Gaussian distribution with a
HWHM in latitude of 8.8$^\circ$ (solid curve). The final dipole
field is roughly proportional to the BMR size and the sine of the
tilt angle \citep{Baumann04}.

\subsection{Dependence of the scatter in the dipole moment at the end of a cycle on
the strength of the cycle} \label{sec:dependence_cycles}

We have also carried out the above analysis for cycle 14 (the
weakest cycle covered by RGO, 2222 sunspot groups) and for cycle 19
(the strongest cycle, 4648 sunspot groups). The results are shown in
Table \ref{tab:cycles14-19}. For our reference cycle 17 (3579
sunspot groups) the scatter in the tilt angle leads to a standard
deviation of 0.78G in the resulting axial dipole moment at the end
of the cycle. The substantially weaker cycle 14 produces a dipole
field with a standard deviation of 0.81G, and the much stronger
cycle 19 is associated with a standard deviation of 0.76G. The
uncertainty of the dipole moment at the end of a cycle resulting
from the random tilt angle scatter is thus almost unrelated to the
strength of the cycle. This can be understood because stronger
cycles have higher mean latitudes \citep{Solanki08,Jiang11a} where
the scatter introduces less noise into the dipole moment (see
Section~4.4). The cycle dependence of the latitude distribution of
sunspot groups is also presented in Table \ref{tab:cycles14-19},
where we see that the differences between the two cycles decreases
at low latitudes. According to the results in the above two
subsections, only the scatter in the tilt angles of large sunspot
groups at low latitudes has large effects on the uncertainties of
the axial dipole moment.

As shown in Table \ref{tab:cycles14-19}, the average dipole moments
for cycles 14 and 19 are 2.01G and 2.21G, respectively.  The
standard deviations correspond to 40\% and 34\%, respectively, of
these values. The percentages are somewhat higher than the value of
28\% for cycle 17., but this is mainly due to the differing
strengths of the dipole moment at the end of the three cycles rather
than to a change in the standard deviation introduced by the
scatter. In absolute terms, the scatter in the dipole moments for
cycles 14, 17 and 19 are almost the same: 0.8G, 0.78G, and 0.75G,
respectively.

Table \ref{tab:cycles14-19} also shows that the near-equator
emergence dominates the scatter of the dipole moments for both weak
and strong cycles: sunspot groups emerging below 15$^\circ$
contribute most to the standard deviation of the axial dipole
moment. Similar to the results for cycle 17 given in Table
\ref{tab:diff_lati}, the combination of the number of sunspot groups
in a given bin and the latitude dependence of their individual
contributions leads to a maximum of the standard deviation in the
5$^\circ$-10$^\circ$ bin.

\begin{figure}
\begin{center}
\includegraphics[scale=1.0]{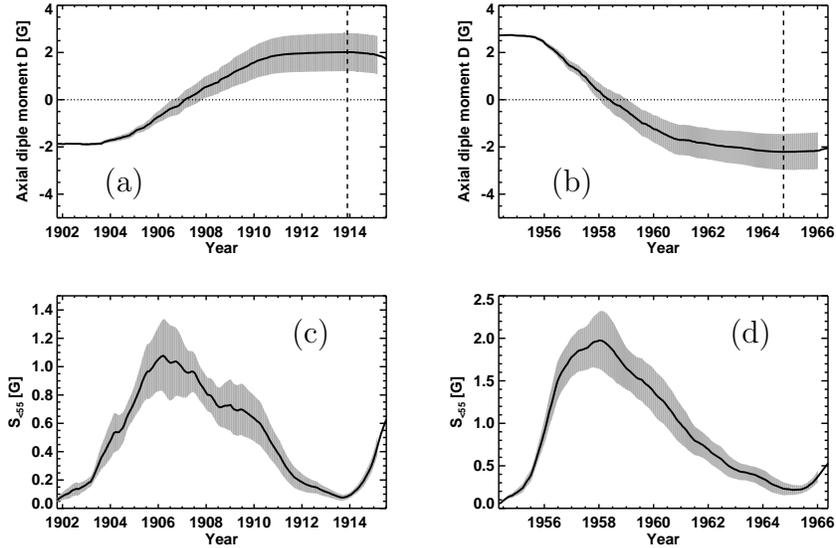}
 \caption{Results for cycle 14 (left panels) and cycle 19 (right panels)
 for simulations including random scatter of the tilt angles. Panels (a)
 and (b): total axial dipole moment; Panels (c) and (d):
mean unsigned latitudinally averaged (signed) flux density  within
$\pm55^\circ$ latitude.}
 \label{fig:cycles14-19}
\end{center}
\end{figure}

Figure \ref{fig:cycles14-19} shows the averages (solid curves) and
the standard deviations (grey shades) of the axial dipole moment and
$S_n$, the mean unsigned latitudinally averaged (signed) flux
densities in the latitudinal range from $-55^\circ$ to $55^\circ$
for cycles 14 and 19.

\section{Summary}
\label{sec:cons}

We have measured the tilt scatter based on the Kodaikanal and Mount
Wilson tilt angle data and studied the effects of this scatter on
the evolution of the solar surface field using the surface flux
transport simulations with flux input based upon the recorded
sunspot groups.

The analysis of the tilt angle data shows that the average tilt
angles have a weak trend to increase with the sunspot group size,
while the standard deviations significantly decrease. The relation
between the standard deviations and the sunspot group size can be
well fitted by a linear logarithmic function.

The simulations including the tilt scatter of the sunspot groups
show that the scatter has a significant effect on the evolution of
the large-scale magnetic field at the solar surface. The
longitudinal averaged magnetic flux at low latitudes is increased
with a more grainy structure generated by sunspot groups with large
tilt angles. On the average, the net unsigned magnetic flux in the
magnetic butterfly diagram at latitudes below 55$^\circ$ is
increased by about 20\%. Including the tilt angle scatter makes the
simulated magnetic butterfly diagram thus better consistent with the
observations: the ratio of the unsigned magnetic flux between the
low and the high latitudes is increased and more poleward surges of
opposite-polarity flux are generated.

The axial dipole moment and the polar fields during solar activity
minimum of cycle 17 may change by about $\pm30$\% (compared to its
mean value) owing to random fluctuations of the tilt angles within
the range indicated by observations. This is consistent with the
results of \citet{Baumann04} for artificial solar cycles. The
standard deviation of the axial dipole field introduced by the
scatter in the tilt angle is almost independent of the strength of
the cycle. The effects of the tilt scatter on the large-scale field
at the solar surface mainly result from large sunspot groups
emerging at low latitudes.

We may conclude from these results that, in the framework of
Babcock-Leighton dynamo models, the random component introduced by
the tilt angle scatter has a significant impact on the variability of the
solar cycle strength. Since the polar fields (or axial dipole moment
of the surface field) that are built up during a cycle represent the
poloidal field source for the following cycle, the fluctuations due
to the tilt angle scatter directly affect the strength of this
cycle.  Even a single big sunspot group with large tilt appearing
near the equator may in this way significantly affect the strength
of the next cycle \citep[cf.][]{Cameron13}. This obviously sets
stringent limits on the predictability of future activity cycles.

\begin{acknowledgements}
We are grateful to the referee for helpful comments on the paper.
J.J. acknowledges the financial support by the National Natural
Science Foundations of China (11173033, 11178005, 11125314,
11221063, 41174153, 2011CB811401) and the Knowledge Innovation
Program of the CAS (KJCX2-EW-T07).
\end{acknowledgements}

\bibliographystyle{apj}
\bibliography{tilt}

\clearpage 

\begin{table}[t]
\caption{Tilt angles of sunspot groups from the Kodaikanal (KOD) and
  Mount Wilson (MWO) records for various ranges of umbral area, $A_U$.
  $N$: number of sunspot groups; $\left\langle \alpha \right\rangle$:
  average tilt angle (in degrees); $\sigma_\alpha$ standard deviation.}
\begin{tabular}{c|ccc|ccc}
\tableline\tableline & & KOD & & & MWO & \\
$A_U$ [$\mu$H]  & $N$ & $\left\langle \alpha \right\rangle$ &
$\sigma_\alpha$ & $N$ & $\left\langle \alpha \right\rangle$ &
$\sigma_\alpha$\\
\tableline
0.0 -- 6.3&  6801  &  4.75$\pm$0.91& 29.97$\pm$0.76&  6660  &  4.63$\pm$0.89& 30.23$\pm$0.75\\
6.3 -- 10.0&  4557  &  4.56$\pm$0.88& 25.63$\pm$0.73&  3939  &  5.25$\pm$0.89& 24.64$\pm$0.73\\
10.0 -- 15.8&  5196  &  4.96$\pm$0.65& 24.02$\pm$0.53&  4686  &  4.94$\pm$0.75& 23.13$\pm$0.61\\
15.8 -- 25.1&  5054  &  5.04$\pm$0.69& 20.78$\pm$0.56&  4758  &  5.36$\pm$0.57& 19.89$\pm$0.46\\
25.1 -- 39.8&  4014  &  5.11$\pm$0.53& 18.16$\pm$0.44&  3822  &  5.51$\pm$0.57& 17.17$\pm$0.47\\
39.8 -- 63.1&  2692  &  6.18$\pm$0.47& 15.55$\pm$0.39&  2583  &  5.14$\pm$0.48& 14.80$\pm$0.39\\
63.1 -- 100.0&  1469  &  5.59$\pm$0.45& 14.94$\pm$0.37&  1293  &  5.24$\pm$0.32& 13.06$\pm$0.26\\
100.0 -- 158.5&   532  &  6.78$\pm$0.38& 10.24$\pm$0.31&   400  &  5.06$\pm$0.43& 10.57$\pm$0.35\\
158.5 -- 251.2&   131  &  6.53$\pm$0.68&  9.55$\pm$0.55&    84  &  5.63$\pm$0.66&  8.29$\pm$0.54\\
251.2 -- max  &    30  &  2.62$\pm$2.50& 14.67$\pm$2.04&    20  &
4.26$\pm$1.02&  6.89$\pm$0.83\\\hline\hline \label{tab:tilts_obs}
\end{tabular}
\end{table}

\begin{table}[t]
\begin{center}
\caption{Standard deviation of the tilt angle scatter of sunspot groups from the
Kodaikanal (KOD) and Mount Wilson (MWO) records for various latitudinal ranges.}
\begin{tabular}{ccccc}
\tableline\tableline
       &0.0$^\circ$ -- 10.0$^\circ$ &  10.0$^\circ$ -- 15.0$^\circ$  &  15.0$^\circ$ -- 20.0$^\circ$ &  20.0$^\circ$ -- max \\
\tableline
KOD    & 30.8       &  29.1        & 30.2   &     31.1\\
MWO    & 30.0       &  29.6        & 30.2   &     30.9\\
\tableline\tableline \label{tab:lat}
\end{tabular}
\end{center}
\end{table}

\begin{table}[t]
\begin{center}
\caption{Maximum values (in G) of various quantities from simulation
runs without and with tilt angle scatter from 50 random
realizations. $D$: axial dipole moment; $B_{\rm NP}$ and $B_{\rm
SP}$: north and south polar field; $S_{<55}$: mean flux of the
magnetic butterfly diagram among $\pm55^\circ$ latitudes. For
definitions, see Eqs.~(\ref{eq:polar}) to (\ref{eq:bavg55}). }
\begin{tabular}{ccc}
\tableline\tableline
 & without scatter & with scatter\\
\tableline
$|D|$ & $2.68$ & $2.73\pm0.78$ \\
$|B_{\rm NP}|$ & $9.46$ & $8.82\pm3.05$\\
$|B_{\rm SP}|$ & $10.82$ & $12.02\pm3.18$\\
$S_{<55}$ & $1.27$ & $1.50\pm0.18$\\
\tableline\tableline
\label{tab:comparison}
\end{tabular}
\end{center}
\end{table}

\begin{table}[t]
\begin{center}
\caption{Results (in G) for tilt angle scatter restricted to sunspot
groups within a given range of umbral area, $A_U$ (in $\mu$H). $N$
denotes the number of the sunspot groups in each bin.}
\begin{tabular}{cccccc}
\tableline\tableline
$A_U$ range & 0 -- 28 & 28 -- 56 & 56 -- 94 & 94 -- 160 & 160 -- max \\
\tableline
$|D|$ & $2.67\pm0.16$ & $2.75\pm0.24$ & $2.69\pm0.33$ & $2.69\pm0.45$ & $2.81\pm0.48$ \\
$|B_{\rm NP}|$ & $8.73\pm0.64$ & $8.98\pm0.95$ & $8.84\pm1.27$ & $8.76\pm1.83$ & $9.14\pm1.74$\\
$|B_{\rm SP}|$ & $11.49\pm0.57$ & $11.83\pm0.94$ & $11.53\pm1.31$ & $11.61\pm1.72$ & $11.99\pm2.01$\\
$N$ & 2496 & 514 & 294 & 179 & 84 \\
\tableline\tableline
\label{tab:diff_size}
\end{tabular}
\end{center}
\end{table}

\begin{table}[t]
\begin{center}
\caption{Results (in G) for tilt angle scatter restricted to sunspot groups
emerging within a given latitude range.}
\begin{tabular}{cccccc}
\tableline\tableline
$|\lambda|$ range & 25$^\circ$ -- max & 15$^\circ$ -- 25$^\circ$ & 10$^\circ$ -- 15$^\circ$ & 5$^\circ$ -- 10$^\circ$ & 0$^\circ$ -- 5$^\circ$ \\
\tableline
$|D|$ & $2.72\pm0.01$ & $2.72\pm0.08$ & $2.71\pm0.36$ & $2.74\pm0.57$ & $2.71\pm0.35$\\
$|B_{\rm NP}|$ & $8.91\pm0.07$ & $8.95\pm0.39$ & $8.86\pm1.42$ & $8.84\pm2.20$ & $8.89\pm1.26$\\
$|B_{\rm SP}|$ & $11.63\pm0.02$ & $11.66\pm0.37$ & $11.68\pm1.67$ & $11.81\pm2.20$ & $11.66\pm1.20$\\
$N$ & 436 & 1249 & 932 & 729 & 223 \\
\tableline\tableline \label{tab:diff_lati}
\end{tabular}
\end{center}
\end{table}

\begin{table}[t]
\begin{center}
\caption{Results for cycles 14 and 19 with the tilt angle scatter
restricted to sunspot groups emerging within a given latitude range.
$D_{14}$ and $D_{19}$: dipole moments; $N_{14}$ and $N_{19}$:
numbers of the sunspot groups in each bin.}
\begin{tabular}{ccccccc}
\tableline\tableline
$|\lambda|$ range & 0$^\circ$ -- max & 25$^\circ$ -- max & 15$^\circ$ -- 25$^\circ$ & 10$^\circ$ -- 15$^\circ$ & 5$^\circ$ -- 10$^\circ$ & 0$^\circ$ -- 5$^\circ$ \\
\tableline
$|D_{14}|$ & $2.01\pm0.81$ & $2.09\pm0.003$ & $2.09\pm0.06$ & $2.11\pm0.24$ & $2.09\pm0.60$ & $2.01\pm0.37$\\
$N_{14}$ & $2222$ & 73 & 770 & 648 & 557 & 174 \\
$|D_{19}|$ & $2.21\pm0.76$ & $2.27\pm0.004$ & $2.28\pm0.13$ & $2.20\pm0.35$ & $2.27\pm0.63$ & $2.27\pm0.43$\\
$N_{19}$ & $4648$ & 789 & 1719 & 1095 & 764 & 281 \\
\tableline\tableline \label{tab:cycles14-19}
\end{tabular}
\end{center}
\end{table}


\end{document}